\documentclass[twocolumn,showpacs,prb,floatfix]{revtex4}
\usepackage{graphicx}
\usepackage{latexsym}

\begin{document}

\title{Quantized conductance in an AlAs 2D electron system quantum point contact }
\author{O. Gunawan}
\author{B. Habib}
\author{E. P. De Poortere}
\altaffiliation{Current address: Department of Physics, Columbia university, New York, NY 10027}
\author{M. Shayegan}
\address{Department of Electrical Engineering, Princeton University, Princeton, New Jersey 08544}
\date{\today}

\begin{abstract}
We report experimental results on a quantum point contact (QPC) device formed in a wide AlAs
quantum well where the two-dimensional electrons occupy two in-plane valleys with elliptical Fermi
contours. To probe the closely-spaced, one-dimensional electric subbands, we fabricated a point
contact device defined by shallow-etching and a top gate that covers the entire device. The
conductance versus top gate bias trace shows a series of weak plateaus at integer multiples of
$2e^2/h$, indicating a broken valley degeneracy in the QPC and implying the potential use of QPC
as a simple "valley filter" device.  A model is presented to describe the quantized energy levels
and the role of the in-plane valleys in the transport. We also observe a well-developed
conductance plateau near $0.7$$\times$$2e^2/h$ which may reflect the strong electron-electron
interaction in the system.
\end{abstract}
\maketitle

\begin{figure}
\includegraphics[width=80mm]{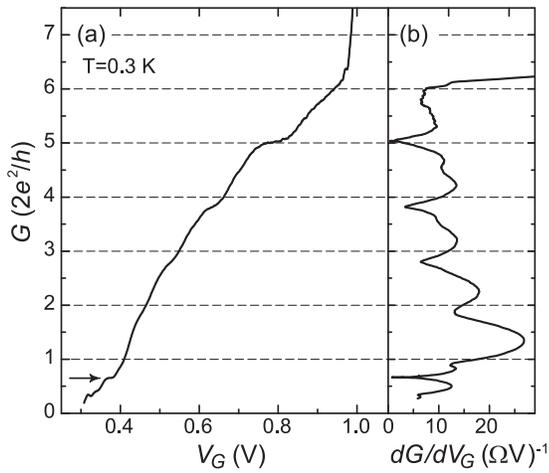} \caption{
(a) Conductance vs gate voltage. The arrow indicates the "0.7 structure".  (b) The
transconductance $dG/dV_{G}$. The minima, nearly periodic in $2e^2/h$ (except for the "0.7
structure"), indicate developing plateaus in the conductance trace. } \label{FigGTrace}
\end{figure}

\section{INTRODUCTION}

The quantum point contact (QPC) is a prime example of a mesoscopic structure that demonstrates a
striking quantum mechanical ballistic transport phenomenon. It exhibits a quantization of
conductance in units of $G_0$$=$$2e^2/h$ where the factor of 2 accounts for spin
degeneracy.\cite{WeesPRL88,WharamJPC88}  This effect has been observed in many two-dimensional
electron systems (2DESs) with mean-free-path longer than the QPC channel length. It arises as a
consequence of the quantization of transverse momentum and full transmission of the
one-dimensional modes in the constriction, reminiscent of quantum Hall effect where the absence of
backscattering leads to quantized plateaus in the Hall resistance. Quantized conductance in QPCs
has been observed and studied in many 2D carrier systems including electrons in GaAs,
\cite{WeesPRL88,WharamJPC88} SiGe,\cite{TobbenSST95,WieserAPL02} GaN,\cite{ChouAPL05} InSb,
\cite{GoelPhysE05} and holes in GaAs.\cite{RokhinsonAPL06,DanneauAPL06}

Here we report transport measurements of a QPC device realized in a 2DES confined to an 11
nm-wide, modulation-doped  AlAs quantum well grown on a GaAs (001) substrate. The novelty of this
system is that the 2D electrons occupy conduction band valleys located at the X points of the
Brillouin zone. There are two anisotropic, elliptical Fermi contours, each characterized by a
heavy longitudinal mass $m_l$$=$$1.1 m_0$ and a light transverse mass $m_t$$=$$0.2 m_0$, where
$m_0$ is the electron mass in vacuum.\cite{AlAsWideQW} Our QPC device is fabricated via
shallow-etching and depositing a top gate that covers the entire sample and controls the density
in both the 2D reservoir and the QPC channel. As outlined in the abstract, the conductance vs gate
bias trace of this device exhibits several interesting features, some of which are puzzling. In
the next section, we summarize these observations and briefly describe their origin; the rest of
the paper is devoted to the details of the device and a quantitative explanation of conductance vs
gate bias data.

\section{QUANTIZED CONDUCTANCE}
Figure~\ref{FigGTrace} presents our main finding. It shows the conductance ($G$) trace of the QPC
device measured at $T$=0.3 K as a function of the applied gate voltage ($V_G$) and the
corresponding transconductance trace ($dG/dV_G$) that accentuates the features in $G$. Three
features of the data are noteworthy. First, we observe developing quantized conductance steps near
integer multiples ($N$) of $2e^2/h$ up to $N=5$; these steps are exhibited more clearly in the
transconductance plot of Fig.~\ref{FigGTrace}(b).\cite{QPCBG} Second, the $2e^2/h$ conductance
plateaus seem to get stronger monotonically at higher $N$. Third, we observe a particularly strong
plateau near 0.7$\times 2e^2/h$.

Perhaps the most puzzling feature of Fig.~\ref{FigGTrace} data is that, even though the
system is expected to have a two-fold valley degeneracy, we do not observe quantized
steps at integer multiples of $4e^2/h$ but instead at multiples of $2e^2/h$. As
we discuss in detail in the remainder of the paper, this happens because the mass
anisotropy breaks the valley degeneracy of the QPC subband energy levels. The valley with
higher mass in the QPC (lateral) confinement direction has lower energy and therefore
dominates the low-lying subband states. Moreover a small but finite residual valley
splitting present in the system further helps lift the valley degeneracy.

\begin{figure}
\includegraphics[width=80mm]{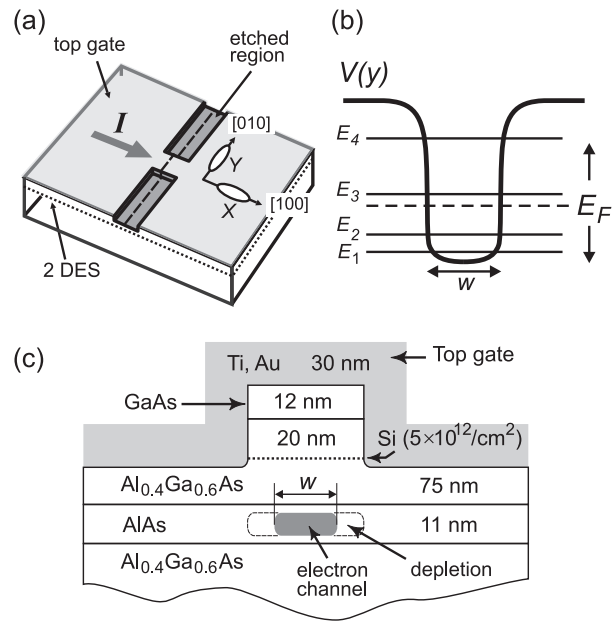} \caption{
(a) Device schematic of the shallow-etched QPC. The $k$-space orientation of the
in-plane valleys labeled as $X$ and $Y$ is also shown. (b) Confinement potential along
the dashed-line in (a). The Fermi energy varies with the gate bias voltage. (c) Device
cross-section along the dashed line in (a). } \label{FigQPCSchematic}
\end{figure}

Next, the overall weakness of the observed plateaus in Fig.~\ref{FigGTrace} imply that the QPC
energy subband spacings are comparable to $k_BT$, where $k_B$ is the Boltzmann constant and $T$ is
the system temperature. This is not surprising, considering the rather large effective mass of
electrons (compared to, e.g., GaAs 2D electrons) which leads to small subband spacings. The fact
that the conductance plateaus get stronger monotonically at higher conductance steps
[Fig.~\ref{FigGTrace}(b)], on the other hand, is somewhat puzzling. This distinct feature is not
observed in conventional surface split-gate QPCs.\cite{WeesPRL88,WharamJPC88}. We attribute it
mainly to two characteristics of our device. First, since our QPC constriction is defined using
shallow-etching and gating (in contrast to only gating using split-gate), we expect the lateral
confinement potential to be strong,\cite{KristensenJAP98, KristensenPRB00} resembling a square
well potential (Fig.~\ref{FigQPCSchematic}). In such a potential, the energy subband spacing
$\Delta E_{N+1,N}$ increases at higher $N$, therefore at a given temperature, the subband spacing
will be better resolved.  Second, the electron mean-free-path in our device is comparable to the
length of the channel. As we increase the gate voltage, the density in the QPC channel increases,
resulting in higher electron mobility and longer electron mean-free-path.\cite{PoortereAPL02} This
results in a better transmission across the QPC and thus a better plateau formation at higher
$V_G$. We note that, a monotonic increase in strength of $2e^2/h$ plateaus suggests that the
quantized energy levels originate from a single valley. Any accidental degeneracy between $X$ and
$Y$ valley 1D energy levels in the QPC would lead to non-monotonic energy level spacings and would
cause a deterioration or missing of conductance plateaus, in contrast to the monotonic increase in
plateau strength as observed here.

Finally, the conductance plateau we observe near 0.7$\times 2e^2/h$ is particularly strong
compared to the other plateaus in our QPC device or the "0.7 structure" plateaus reported for QPCs
fabricated using other 2D systems. The origin of the "0.7 structure" is still unknown although
there is a general consensus that, unlike the integer plateaus, it may be caused by
electron-electron interaction.
\cite{ThomasPRL96,KristensenPRB00,CronenwettPRL02,ReillyPRL02,GrahamPRL03,RochePRL04} Our
observation is consistent with such interpretation as we indeed do expect interaction to be strong
in our system because of the heavy electron effective mass.

\section{DEVICE STRUCTURE}
We used a modulation-doped AlAs quantum well (QW) grown by molecular beam epitaxy on a (001) GaAs
substrate. The well is 11 nm wide, resides $\simeq$110 nm below the surface and is flanked by
undoped and Si $\delta$-doped layers of Al$_{0.4}$Ga$_{0.6}$As [Fig.~\ref{FigQPCSchematic}(c)].
\cite{PoortereAPL02} The QPC is fabricated on a standard Hall bar device, defined by optical
lithography.

In most QPC devices, the channel is defined by a pair of surface metal split gates to control the
channel's electrical width using an applied voltage bias to the gates.\cite{WeesPRL88,
WharamJPC88} In such a device the width of the confinement potential changes considerably while
the Fermi level ($E_F$) in the 2DES reservoir remains constant. In a system with large electron
effective mass such as in AlAs 2DES, the energy level spacings are expected to be very small,
rendering the observation of quantized conductance challenging. We tried a number of split-gate
structures with no success in observing quantized conductance steps. Experiments on high quality
AlAs quantum wires fabricated via the cleaved-edge-overgrowth technique, have not yielded clear
quantized plateaus either.\cite{MoserAPL05} We therefore took an alternative approach and defined
the QPC constriction by shallow-etching to introduce a strong QPC confinement,
\cite{KristensenJAP98, KristensenPRB00} and then covered the entire device by a Ti/Au top gate
that controls the 2DES density (and thus $E_F$) in the reservoir and in the QPC constriction
[Fig.~\ref{FigQPCSchematic}(a)]. As we demonstrate later in the paper, this device structure
allows us to directly probe the energy spacing between the QPC quantized energy levels. One
disadvantage is that the background resistance, from the series resistance of the 2DES reservoir
regions flanking the point contact, varies with the gate voltage; however, this background
correction is found to be rather negligible as shown in Fig.~\ref{FigGTrace}.\cite{QPCBG}

Our QPC constriction was defined by electron beam lithography to be 300 nm wide and 500 nm long
with the transport direction aligned along [100] as shown in Fig.~\ref{FigQPCSchematic}(a). Due to
carrier depletion near the boundary of the constriction, however, the electrical channel width
($w$) is expected to be smaller. We performed wet-etching using H$_2$SO$_4$:H$_2$O$_2$:H$_2$0
(1:8:160) solution to a depth of 65 nm to remove the Si dopant layer, thus completely depleting
the electrons under the etched region [Fig.~\ref{FigQPCSchematic}(c)]. Using illumination and
top/back gate biasing,\cite{PoorterePRB03} we were able to vary the 2DES density $n_{2D}$ between
$2.4$ to $4.5$$\times$$10^{15}$ /m$^2$ with a low temperature mobility of $\sim$15 m$^2$/Vs,
implying a transport mean-free-path of $\sim$1 $\mu$m.  The measurements were done in a $^3$He
cryostat system with 0.3 K base temperature and using standard phase sensitive lock-in technique
in a four-wire configuration.

\section{ANALYSIS AND DISCUSSION}

To gain insight into the transport in the QPC and especially to understand the role of the two
in-plane valleys, we present a simple model to describe the QPC energy levels by taking into
account the QPC geometry and the electron system parameters. \textit{Our goal is to accurately
explain the values of $V_G$ at which we observe the conductance plateaus.} We start with an
infinite square well model and a constant QPC channel width. However, as we will show, this
description is unsatisfactory for our system, and thus later on we refine the model by treating
the channel width as a gate voltage-dependent variable. We also corroborate this analysis with the
channel width deduced from our low-field magnetoresistance data.

\subsection*{A. Quantum point contact model}

\begin{figure}
\includegraphics[width=80mm]{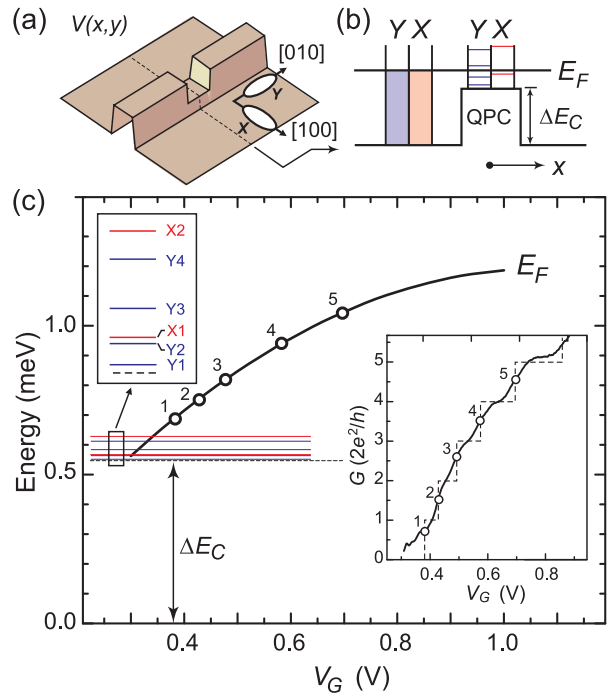} \caption{
(a) Schematic drawing showing the potential landscape surrounding the QPC and the orientations of
the two in-plane valleys $X$ and $Y$. (b) Potential cross-section along the transport direction
[dotted line in (a)]. (c) Dependence of the Fermi energy on to the gate voltage. The circles are
the expected level crossings. Right inset: The conductance vs $V_G$ trace together with an
idealized conductance step model to determine the level crossings (see text). Left inset: The QPC
energy levels assuming a fixed channel width of $w=300$ nm. It is apparent that the energy levels
do not fit the level crossings. } \label{FigQPCModelX}
\end{figure}

Figures~\ref{FigQPCModelX}(a) and (b) present a model for the potential landscape around the QPC.
In our device, as we increase the gate voltage, $E_F$ increases and crosses the quantized energy
levels in the QPC, leading to quantized conductance steps. The Fermi energy can be obtained from
the density ($n_{2D}$) in the 2D reservoirs on the two sides of the QPC. This density can be
determined from Shubnikov-de Haas oscillations and Hall resistance measurements as will be
detailed in the next section (Figs.~\ref{FigQPCMR} and~\ref{FigFFT}). We start with the simple
assumption that there are two in-plane valleys $X$ and $Y$ with equal population in the 2D
reservoirs participating in transport; $X$ and $Y$ refer to valleys with their major axes along
[100] and [010] directions (see Figs.~\ref{FigQPCSchematic} and~\ref{FigQPCModelX}). Given
$n_{2D}$, we can calculate the Fermi energy $E_F=n_{2D}/\rho$ where $\rho=2m^*/\pi \hbar^2$ is the
spin- and valley-degenerate 2D density of states; we used $m^*=\sqrt{m_l m_t}$. This $E_F$ is
plotted as a function of $V_G$ in Fig.~\ref{FigQPCModelX}(c).

From the conductance trace in Fig.~\ref{FigGTrace} we can estimate the gate voltages where $E_F$
crosses the 1D subbands in the QPC. Since the plateaus are not sharply defined in our experiment
we cannot pinpoint exactly where these crossing points are. However the kinks at every $2e^2/h$ in
the conductance trace are clear. These kinks occur when $E_F$ lies in the middle of an energy gap.
Thus the level crossings occur approximately at gate voltages half-way in between successive gate
voltages where the kinks appear. Using this approximation we construct an idealized conductance
steps trace shown by dashed lines in the inset of Fig.~\ref{FigQPCModelX}(c). The positions where
the level crossings occur are marked as circles (labeled with their subband indices $N=1$ to 5).

Figure~\ref{FigQPCModelX}(b) assumes that the QPC has an elevated bottom potential, with an offset
$\Delta E_C$, with respect to the 2DES reservoir potential. This is to model the fact that the QPC
has a pinch-off voltage $V_P=0.3$ V (Fig.~\ref{FigGTrace}) even though the density in the 2D
reservoir is non-zero at this voltage. We take $\Delta E_C$ to be equal to $E_F$ at $V_G = V_P$.

In the QPC channel, the $X$ and $Y$ valleys have a different set of quantized energy levels as
shown in Figure~\ref{FigQPCModelX}(b). This happens because, thanks to the different valley
orientations, the electrons in the $X$ and $Y$ valleys have different effective masses along the
lateral confinement direction. This mass anisotropy breaks the degeneracy of the quantized levels
in the channel. Since the confinement potential is strong and quantized energies are small due to
heavy electron effective mass, we can estimate the 1D subbands' energy levels using an infinite
square well model: $E_N=N^2\hbar^2\pi^2/2m_y^*w^2$ where $m_y^*$ is the electron mass along the
QPC confinement direction; $m_y^*$ is equal to $m_l$ ($m_t$) for the $Y$ ($X$) valley. Thus the
ground state energy for valley $Y$ is lower than that of $X$ by a factor of $m_l/m_t=5.5$.
Finally, given the channel width of $w$=300 nm, we calculated the first few subband levels for
both $X$ and $Y$ valleys and plot them in Fig.~\ref{FigQPCModelX}(c). It is apparent that the
energy levels do not agree with the expected level crossings. This points to the inadequacy of our
simple model and demands further refinement, which we describe in the next section.

\subsection*{B. Determination of QPC channel density and width}

\begin{figure}
\includegraphics[scale=1]{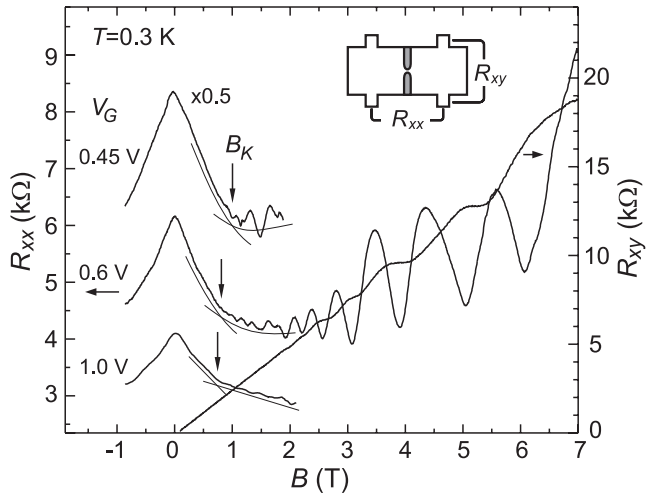} \caption{
Magnetoresistance traces for the QPC device. Inset shows the contact geometry for these traces.
The kink in the $R_{xx}$ magnetoresistance trace, (marked by the arrow $B_K$), signals the field
where the cyclotron orbit fits into the width of the QPC channel. For $V_G=0.6$ V, a Hall
resistance ($R_{xy}$) trace is also shown. } \label{FigQPCMR}
\end{figure}

\begin{figure}
\includegraphics[scale=1]{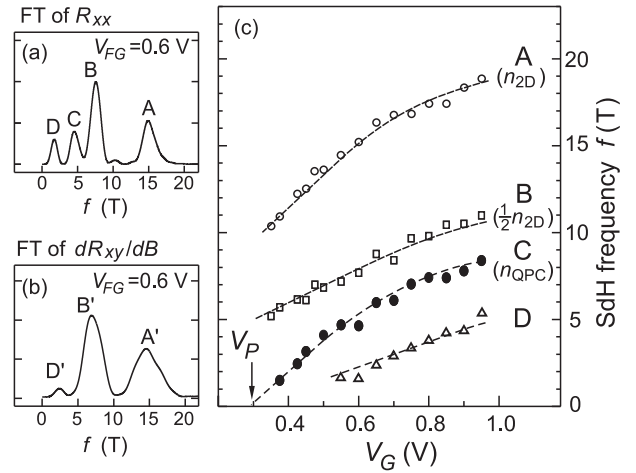} \caption{
(a) Fourier transform (FT) spectrum of $R_{xx}$ representing density components in the 2D
reservoir and the QPC. (b) Fourier transform spectrum of $dR_{xy}/dB$ representing density
components in the 2D reservoir alone. (c) Summary of the Fourier spectra peaks vs $V_G$.}
\label{FigFFT}
\end{figure}

Since the constriction is defined by etching, the removal of the dopant layer, together with the
Fermi level pinning effect at the exposed surface, induces depletion regions that narrow the QPC
channel. Simple theoretical estimates, confirmed by experimental results, indicate that the
depletion width decreases with increasing 2DES density.\cite{ChoiAPL87, TakagakiSST90} In
addition, as seen in Fig.~\ref{FigQPCSchematic}(c), the top gate covering the shallow-etched
region comes to close proximity of the QPC channel (somewhat similar to a split-gate device) so
that changing the gate bias may also influence the channel width electrostatically. Therefore, we
expect that the channel width $w$ would increase at higher $V_G$ where the density is higher.

It is possible to estimate the channel width from the QPC magnetoresistance
(MR).\cite{vanHoutenPRB88} Figure~\ref{FigQPCMR} shows the longitudinal MR ($R_{xx}$) traces
across the QPC and the transverse MR ($R_{xy}$) measured in one of the reservoir regions (see
Fig.~\ref{FigQPCMR} inset). The $R_{xx}$ traces exhibit strong negative MR at low field. This
negative MR arises from the suppression of the constriction resistance by the magnetic field in
the ballistic regime. As the magnetic field is increased, the electron backscattering rate is
reduced and a larger fraction of the edge states are transmitted through the channel, resulting in
a smaller resistance.\cite{vanHoutenPRB88} This behavior persists up to a magnetic field $B_K$ at
which the classical cyclotron diameter equals the channel width. At $B_K$ there is a marked change
in the MR slope, appearing as a "kink" in the trace (Fig.~\ref{FigQPCMR}) that can be used to
estimate the width of the QPC:
\begin{eqnarray}
 w =2\hbar k_F /e B_K
 \label{wBk}
\end{eqnarray}
where $k_F$ is the Fermi wavevector perpendicular to the width of the QPC.\cite{vanHoutenPRB88}

In our QPC there are two possible in-plane valleys, each with an elliptical cyclotron orbit, that
could participate in the transport. For the $X$ and $Y$ valleys, the Fermi wavevectors along the
transport direction (perpendicular to the channel width) can be written as:\cite{GunawanPRL04}
\begin{eqnarray}
  k^2_{F,X}=2\pi n_{QPC,X} \sqrt{m_l/m_t}\\
  \label{kF1}
  k^2_{F,Y}=2\pi n_{QPC,Y} \sqrt{m_t/m_l}
  \label{kF2}
\end{eqnarray}
where $n_{QPC,X}$ and $n_{QPC,Y}$ are the 2D valley densities in the QPC for the $X$ and $Y$
valleys, respectively. These electron densities in the QPC channel can be determined from the
Fourier spectra of the Shubnikov-de Haas (SdH) oscillations as follows.

Since $R_{xx}$ is measured across the QPC (Fig.~\ref{FigQPCMR}), it contains SdH oscillation
contributions from both the 2D reservoirs and the QPC region. \cite{BerggrenPRL86} Ideally, we
could obtain density information in the 2D reservoirs from MR measurements in a plain 2D region
adjacent to the QPC. Unfortunately, there are no contacts available to measure such a region in
our device. Thus, to obtain the density in the 2D reservoirs alone, we deduce the SdH oscillations
from $dR_{xy}/dB$ that represents the "longitudinal MR" following the empirical resistance
rule.\cite{PanPRL05} It has been shown that the Hall resistance is unaffected by the presence of
the QPC constriction. \cite{vanHoutenPRB88}

The Fourier spectra of $R_{xx}$ and $dR_{xy}/dB$ are shown in Figs.~\ref{FigFFT}(a) and (b). We
observe three common frequency components labeled as A, B and D in Fig.~\ref{FigFFT}(a) [A', B',
and D' in Fig.~\ref{FigFFT}(b)] in the Fourier spectra of $R_{xx}$ and $dR_{xy}/dB$ traces. These
frequency components are associated with total density (A) and half total density (B) in the 2D
reservoir region\cite{ShkolnikovPRL02} (We will shortly return to peak D in the next section). We
assign the frequency component C, which appears only in $R_{xx}$ Fourier spectrum
[Fig.~\ref{FigFFT}(b)], to the total density in the QPC channel. Our justification for this
assignment is two-fold. First, the frequency component C does not appear in the SdH spectrum of
$dR_{xy}/dB$ that represents SdH oscillations from the 2D reservoir alone. Second, this frequency
component extrapolates to zero at $V_G$ = 0.3 V [Fig.~\ref{FigFFT}(c)] marking the pinch-off
voltage $V_P$ for the QPC, consistent with the onset of the conductance trace
(Fig.~\ref{FigGTrace}). Therefore, from the peak C positions, we can determine the electron
densities in the QPC channel using: $n_{QPC}=f_C$$\times$$e/h$.

\begin{figure}
\includegraphics[width=80mm]{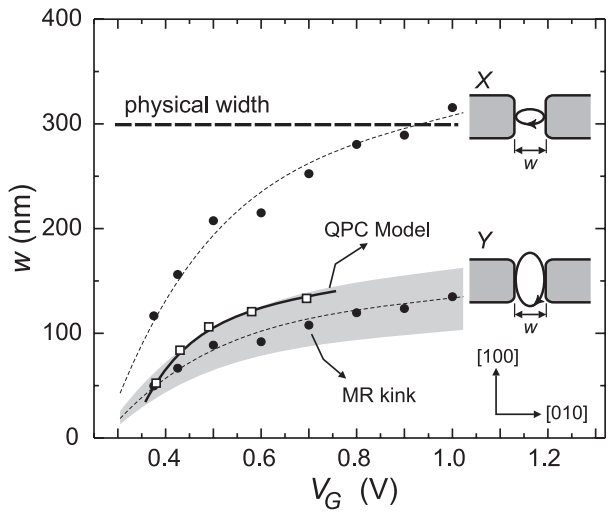} \caption{
QPC channel electrical width ($w$), deduced from the kink in the magnetoresistance, for two
possible cases where either valley $X$ or $Y$ is occupied; these cases are schematically
illustrated on the right. Gray area denotes the error band for $w$ from the MR kink data. The open
squares and solid line indicate $w$ determined from the QPC model (see text).} \label{FigQPCWidth}
\end{figure}

Assuming that only one valley is present in the channel, there are two possible cyclotron orbit
orientations and the width $w$ can be calculated for each case based on Eqs.~\ref{wBk}-\ref{kF2}
as shown in Fig.~\ref{FigQPCWidth}. Using a scanning electron microscope, we confirmed that the
QPC geometrical width is indeed 300 nm, consistent with the width intended in our electron-beam
lithography process. This implies that the width deduced for $X$ valley is not realistic since it
exceeds 300 nm at high gate voltage. We conclude that it is the $Y$ valley that is occupied in the
QPC and therefore dominates the transport across the QPC channel. This gives channel widths
ranging from 50 to 130 nm (Fig.~\ref{FigQPCWidth}), implying depletion widths of 85 to 125 nm on
each side of the channel wall. These are reasonable values for depletion widths in this type of
structure. \cite{ChoiAPL87}

\subsection*{C. Quantum point contact model - revisited}
\begin{figure}
\includegraphics[width=80mm]{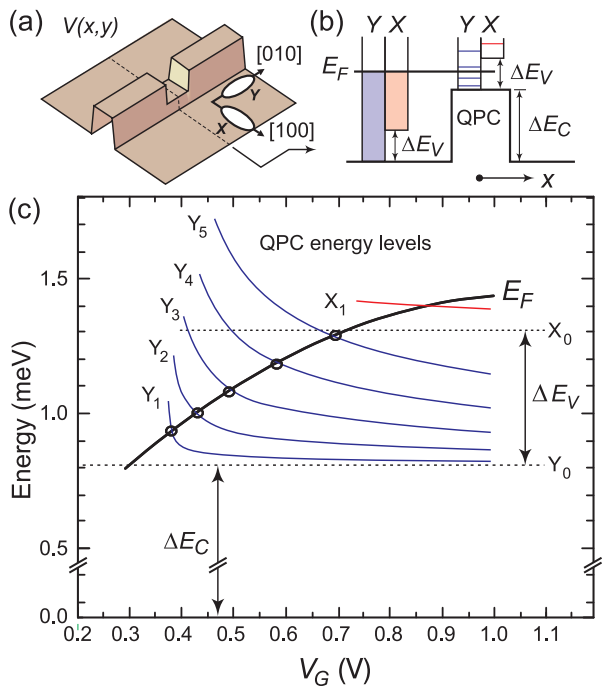} \caption{
(a) Potential landscape surrounding the QPC. (b) Potential cross-section along the transport
direction. A finite residual valley splitting ($\Delta E_V$) in the 2D reservoir is shown. In the
QPC this valley splitting is enough to completely depopulate the $X$ valley. (c) A revised QPC
energy level model with variable channel width. The model fits the QPC energy levels to all
observed level crossings.
 } \label{FigQPCModelOk}
\end{figure}

Having established that the QPC width increases with increasing gate voltage, we can now refine
our QPC model (Fig.~\ref{FigQPCModelOk}). As mentioned in Section II, the monotonic increase in
plateau strength of the conductance trace with increasing $V_G$ suggests that the transport in the
QPC arises from a single valley. This situation requires a residual valley splitting that further
lifts the valley degeneracy and depopulates one of the valleys. Indeed from the Fourier spectrum
of $dR_{xy}/dB$ in Fig.~\ref{FigFFT}(b), where the spectrum represents SdH oscillation components
arising from the 2D reservoir only, the half valley density peak D (at $\sim$2.5 T) and half total
density peak B (at $\sim$7 T) indicate a valley population imbalance. This imbalance corresponds
to a valley splitting of $\Delta E_V\simeq0.5$ meV. In the presence of such a residual valley
splitting, the Fermi energy of the majority valley is given as $E_F=n_{2D}/\rho+\Delta E_{V}/2$,
as plotted in Fig.~\ref{FigQPCModelOk}(c).

The presence of a residual valley splitting in our sample is not unusual. It normally arises from
a built-in residual strain introduced during sample cooldown. Furthermore, this valley splitting
could be enhanced in the QPC channel. First, it is possible that additional anisotropic strain may
be introduced during the fabrication process steps such as etching and gate evaporation. Second,
the additional confinement and lower electron density in the QPC region could lead to stronger
exchange interaction and make the system more easily valley polarized.\cite{GunawanEsMs06} Thus it
is not unlikely that the transport in the QPC channel is valley polarized.

Therefore, in the refined model, we assume that there is only one valley that dominates the
transport in the QPC, in this case the $Y$ valley as suggested from the previous section and
Fig.~\ref{FigQPCWidth}. Using an infinite square well model with channel width $w$ as an
adjustable parameter, we can calculate the 1D energy subbands to fit the expected level crossings
as shown in Fig.~\ref{FigQPCModelOk}(c). We assume the width $w$ is a monotonic and smooth
function of the gate voltage. As we increase $E_F$, the QPC channel gets wider and the quantized
energy levels drop. We obtain the gate voltage-dependent width $w(V_G)$ as the fitting parameter
and plot $w$ as a solid curve in Fig.~\ref{FigQPCWidth} for comparison. We find that the widths
deduced from the QPC model fall fairly close, within the error band, to the widths deduced from
the negative MR kink. This self consistency indicates that indeed it is the $Y$ valley that
dominates the transport in the QPC, while the $X$ valley in the QPC is depopulated due to a
residual valley splitting.

The resulting energy level spacings between successive quantized energy levels are $\sim$$0.1$
meV, very small compared to typical subband spacing of most GaAs QPC devices ($\sim$5 to 20 meV),
yet still somewhat larger than the thermal energy $\sim$$26$ $\mu$eV at $T$ = 0.3 K. This is
consistent with the fact that we observe only rather weak developing, quantized conductance
plateaus.

We would like to point out that we can interpret the data of Fig.~\ref{FigQPCModelOk} with the $X$
valley instead of $Y$ as the source of the quantized energy levels, and $m_t$ as the relevant
effective mass along the confinement potential. In order to yield the same energy levels, such an
interpretation requires a larger channel width, whose values fall near (slightly above) curve $X$
in Fig.~\ref{FigQPCWidth}. This happens because the quantized energy levels $E_N$ scale as
$1/m^*w^2$ and, to obtain the same energy levels, the channel width has to be larger by
$\sqrt{m_l/m_t}$. However, as mentioned in section IV.B, the widths deduced for the case of $X$
valley are not realistic as they exceed the physical channel width of 300 nm at high $V_G$.

\section{SOURCE DRAIN BIAS SPECTROSCOPY}

\begin{figure}
\includegraphics[width=75mm]{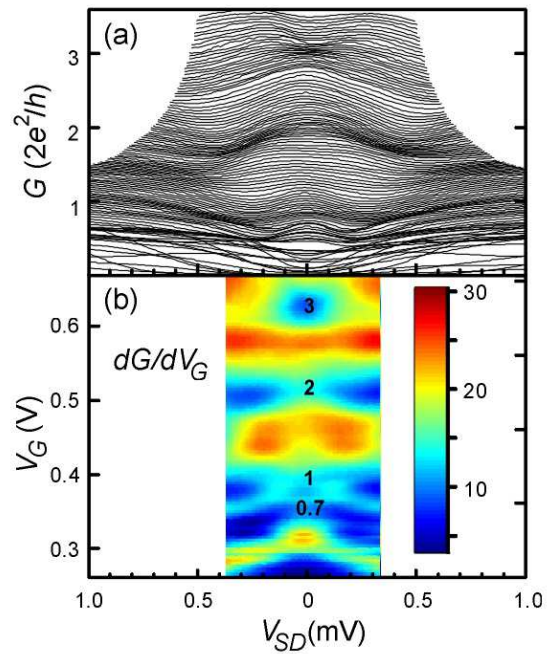} \caption{
(a) Differential conductance $G=dI/dV_{SD}$ traces measured at $T$=0.3 K. Each trace corresponds
to incremental step of 2 mV in $V_G$ starting from 0.3 V at the bottom. (b) The transconductance
($dG/dV_{G}$) spectrum. The bluish (dark) regions correspond to developing plateaus. The numbers
represent the plateaus indices. } \label{FigVSDSpect}
\end{figure}

\begin{figure}
\includegraphics[width=80mm]{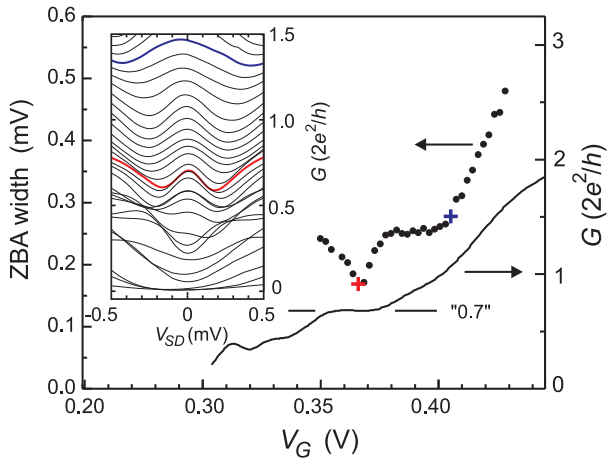} \caption{
Full-width at half-maximum (FWHM) of the ZBA peak as a function of gate voltage. The color data
points (plus signs) are associated with the conductance trace of the same color in the inset. For
reference the conductance trace $G$ vs $V_G$ is replotted. Inset: Differential conductance
$G=dI/dV_{SD}$ at low conductance where the ZBA occurs. } \label{FigZBA}
\end{figure}

Having provided a basic understanding of the QPC conductance data vs $V_G$ in our system, we
proceed to focus on the "0.7 structure". A relevant measurement is the finite source-drain bias
spectroscopy.\cite{CronenwettPRL02,KristensenPRB00,ChouAPL05} We measured the differential
conductance $G=dI/dV_{SD}$ by applying a small AC excitation signal superposed on a DC
source-drain bias ($V_{SD}$). Figure~\ref{FigVSDSpect}(a) shows that the measured source-drain
bias conductance traces exhibit non-linear transport behavior through the QPC. We observe a
bunching of the traces along the $V_{SD}=0$ line that indicate the formation of plateaus at
integer multiples of $2e^2/h$.

These features in the conductance traces can be seen more clearly in the transconductance spectrum
$dG/dV_{G}$ obtained numerically. This spectrum is displayed in Fig.~\ref{FigVSDSpect}(b) where
the blue color indicates minima in $dG/dV_{G}$ that correspond to developing conductance plateaus
at integer multiples of $2e^2/h$. If the plateaus are well developed one can extract the energy
subband spacings in the QPC.\cite{PatelPRB91, KristensenPRB00} Unfortunately, since the plateaus
are weak in our device,  we cannot deduce the energy subband level spacings from this data.

However, one particularly interesting feature of the non-linear conductance measurements, relevant
to the "0.7 structure", is the conductance peak around zero $V_{SD}$, known as the zero bias
anomaly (ZBA).\cite{CronenwettPRL02} This feature can be observed more clearly in the expanded
plot of the $V_{SD}$ bias spectrum as shown in the inset of Fig.~\ref{FigZBA}. The ZBA below the
first $2e^2/h$ plateau has been observed previously in other systems such as in GaAs
\cite{CronenwettPRL02} and GaN \cite{ChouAPL05} QPCs and has been associated with a Kondo-like
correlated state. The width of the ZBA was found to show a dip near the "0.7 structure" and
increase monotonically with increasing gate voltage.\cite{CronenwettPRL02,ChouAPL05} Here, we
observe qualitatively similar behavior (Fig.~\ref{FigZBA}). It was argued in
Ref.~\onlinecite{CronenwettPRL02} that the widths of the ZBA peaks, after the dip, are close to
$2k_B T_K/e$ where $T_K$ is the Kondo temperature of the system. However, we remark on a
difference here. It appears in our data that, for our device, the ZBA persists above the first
quantized plateau ($2e^2/h$) and its width keeps increasing, while in the GaAs system the ZBA
width diverges as the conductance traces collapse to the first plateau. \cite{CronenwettPRL02}
This could simply be related to the fact that our QPC does not show a strong plateau at $2e^2/h$.

\section{SUMMARY AND CONCLUSIONS}
We have studied transport in a QPC device fabricated in a two-dimensional electron system confined
to an 11 nm-wide AlAs QW. The QPC is defined by shallow-etched regions and entirely covered by a
top gate that controls the density in both the 2D reservoirs and the QPC channel. The conductance
trace obtained shows developing quantized conductance plateaus at integer steps of $2e^2/h$. From
the density measurement in the 2D reservoirs and in the QPC channel we construct a simple model
with a gate-voltage dependent channel width to describe the crossings of the Fermi energy and the
QPC quantized energy levels. The QPC model and the channel width determination using the negative
magnetoresistance kink corroboratively indicate that the transport in the QPC channel is dominated
by the $Y$ valley, the valley with higher mass along the QPC lateral confinement direction. This
is a plausible outcome since the lowest QPC subband energy levels should be dominated by the
carriers with higher effective mass along the confinement direction. Our results highlight the
intricacies of understanding quantized conductance in a system with multiple and anisotropic
valleys such as ours. Additionally, we observe a well-developed "0.7 structure" that may reflect
strong electron-electron interaction in this system thanks to its heavier electron mass.

We close by making a few remarks. There has been considerable interest recently in utilizing the
spin degree of freedom to make functional devices (spintronics),\cite{WolfSci01} or to serve as a
qubit for quantum computation.\cite{PettaSci05} The valley degree of freedom may provide an
alternative for such applications.\cite{GunawanEsMs06} The QPC can then be utilized as a simple
valley filter, a potentially important device component, where the QPC confinement breaks the
valley degeneracy, favoring the carriers in one valley and thus leading to valley polarized
transport. In contrast, a spin filter is much more difficult to implement.\cite{FolkSci03}
Furthermore, by simply orienting the QPC transport directions along either [100] or [010], one can
choose which valley species to filter. Alternatively, one could preserve the valley degeneracy in
the QPC by orienting the QPC with its transport direction along [110]. Finally, it is possible to
control the valley populations by applying tunable strain using a piezo-actuator to which the
sample is glued.\cite{ShayeganAPL03}

\section*{ACKNOWLEDGEMENT}
We thank the NSF and ARO for support, and K. Vakili, Y. P. Shkolnikov, D. Goldhaber-Gordon and J.
Shabani for illuminating discussions.


\begin{thebibliography}{33}
\bibitem[{QPC()}]{QPCBG}
\bibinfo{note}{Small discrepancies between the actual plateau positions and
  $N$$\times$$2e^2/h$ levels are present due to background resistances of 100
  to 800 {$\Omega$}, which are typical values for resistance of the 2D
  reservoir regions.}
\end{thebibliography}
\end{document}